\documentclass[twocolumn,showpacs,  preprintnumbers, superscriptaddress, amssymb,pra]{revtex4-1}  

\usepackage{amssymb}
\usepackage{amsmath}
\usepackage{graphicx}
\usepackage{dcolumn}
\usepackage{bm}
\usepackage{verbatim}
\usepackage{array}
\usepackage{epsfig}
\usepackage{amsbsy}	
\usepackage{color}
\usepackage{soul}
\usepackage{url}
\usepackage{hyperref}
\usepackage{natbib}
\usepackage{longtable}

\usepackage[mediumqspace,mediumspace,amssymb,Gray]{SIunits}

\def\rmem#1#2#3{  \left\langle #1 \left\vert \left\vert  #2
                  \right\vert \right\vert #3 \right\rangle   }

\def\normv#1#2{ \left\langle   #1 |  #2 \right\rangle   }

\definecolor{darkgreen}{rgb}{0.0, 0.6, 0.6}

\begin{document}


\preprint{\hfill\parbox[b]{0.3\hsize}{ }}

\title{ Relativistic evaluation of the two-photon decay of the \\ metastable ${1s}^{2} 2s 2p$ $^3\mbox{P}_0$ state in  berylliumlike ions \\with an active-electron model}

%
%

%
\author{Pedro Amaro}
\email{pdamaro@fct.unl.pt}
\affiliation{Laborat\'orio de Instrumenta\c{c}\~ao, Engenharia Biom\'edica e F\'isica da Radia\c{c}\~ao
(LIBPhys-UNL),~Departamento de F\'isica, Faculdade~de~Ci\^{e}ncias~e~Tecnologia,~FCT,~Universidade Nova de Lisboa,~P-2829-516 Caparica, Portugal.}

\author{Filippo Fratini}
\affiliation{Atominstitut, Vienna University of Technology, A-1020 Vienna, Austria}

\author{Laleh Safari}
\affiliation{IST Austria, Am Campus 1, A-3400 Klosterneuburg, Austria.}

\author{Jorge Machado}
\affiliation{Laborat\'orio de Instrumenta\c{c}\~ao, Engenharia Biom\'edica e F\'isica da Radia\c{c}\~ao
(LIBPhys-UNL),~Departamento de F\'isica, Faculdade~de~Ci\^{e}ncias~e~Tecnologia,~FCT,~Universidade Nova de Lisboa,~P-2829-516 Caparica, Portugal.}
\affiliation{Laboratoire Kastler Brossel, \'Ecole Normale Sup\'erieure, CNRS,  
Universit\'e P. et M. Curie -- Paris 6, Case 74; 4, place Jussieu, F-75252 Paris CEDEX 05, France.}

\author{Mauro Guerra}
\affiliation{Laborat\'orio de Instrumenta\c{c}\~ao, Engenharia Biom\'edica e F\'isica da Radia\c{c}\~ao
(LIBPhys-UNL),~Departamento de F\'isica, Faculdade~de~Ci\^{e}ncias~e~Tecnologia,~FCT,~Universidade Nova de Lisboa,~P-2829-516 Caparica, Portugal.}

\author{Paul Indelicato}
\affiliation{Laboratoire Kastler Brossel, \'Ecole Normale Sup\'erieure, CNRS,  
Universit\'e P. et M. Curie -- Paris 6, Case 74; 4, place Jussieu, F-75252 Paris CEDEX 05, France.}

\author{Jos\'e Paulo Santos}
\affiliation{Laborat\'orio de Instrumenta\c{c}\~ao, Engenharia Biom\'edica e F\'isica da Radia\c{c}\~ao
(LIBPhys-UNL),~Departamento de F\'isica, Faculdade~de~Ci\^{e}ncias~e~Tecnologia,~FCT,~Universidade Nova de Lisboa,~P-2829-516 Caparica, Portugal.}

\date{Received: \today  }

\begin{abstract}
The two-photon ${1s}^{2} 2s 2p$ $^3\mbox{P}_0 \rightarrow {1s}^{2} {2s}^2$ $^1\mbox{S}_0$  transition  in berylliumlike ions is theoretically investigated within a full relativistic framework and a second-order perturbation theory.  We focus our analysis on how electron correlation, as well as the negative-energy spectrum can affect the forbidden $E1M1$ decay rate. For this purpose we include the electronic correlation by an effective potential and within an active-electron model. Due to its experimental interest, evaluation of decay rates are performed for berylliumlike xenon and uranium. We find that the negative-energy contribution can be neglected in the present decay rate. On the other hand, if  contributions of electronic correlation are not carefully taken into account, it may change the lifetime of the metastable state by 20\%.  By performing a full-relativistic $jj$-coupling calculation, we found discrepancies for the decay rate of an order of 2 compared to non-relativistic $LS$-coupling calculations, for the selected heavy ions.
\end{abstract}


\maketitle


\section{Introduction}
\label{intro}

 Two-photon decay has been studied several times since it was originally discussed by G\"{o}ppert-Mayer \cite{gop1931}. In low-$Z$ atomic systems, the $2s - 1s$ transitions in hydrogenlike and heliumlike ions occur primarily by two electric dipole photons ($E1E1$), and the respective decay rates provided by theory and experiment are in good agreement.  These works focused not only on the total and energy differential decay rates \cite{sab1959, mas1972, flo1984}, but also on the angular and polarization correlations of the emitted two photons \cite{auu1976, adr1982, kdb1997, fmt2011, sas2014}. Detailed analysis of these two-photon properties have been used to reveal unique information about electron densities in astrophysical plasmas and thermal x-ray sources, as well as highly precise values of physical constants \cite{sjb1999}. The study of two-photon decay in high-$Z$ ions also provided a sensitive tool for exploring the relativistic and quantum electrodynamic (QED) effects that occurs in the strong atomic fields of those systems. As in the case of low-$Z$ ions, predictions for two-photon decay rates are in good agreement with experimental data \cite{god1981, dej1997, tkv2010, dhb1989}.	

Scarce investigations have been performed so far for {\it other} atomic systems with more than two electrons.  In the case of lithiumlike ions, this lack of research might be attributed  to almost all two-photon transitions being in direct competition with dominant allowed (single $E1$) transitions, thus reducing the importance of the former process in practical applications. 
However, this is not the case for berylliumlike ions with zero nuclear spin ($I=0$). Owing to the $0\rightarrow0$ selection rule, the first excited state ${1s}^{2} 2s 2p$ $^3\mbox{P}_0$ is metastable and its transition to the ground state ${1s}^{2} 2s^2$ $^1\mbox{S}_0$ is strictly forbidden for all single-photon multipole modes. The most dominant decay process is a rare two-photon transition with a magnetic dipole mode ($E1M1$) that is very sensitive to relativistic and electronic correlation effects and can have lifetimes from few decades to few minutes, depending on the atomic electromagnetic field of the nucleus. 

Knowledge of metastable decay rates are essential in collision-radiative modelling of astrophysical low-density plasmas that occurs in stellar coronae \cite{tra2014}, thus many studies have been dedicated to the measurement and calculation of higher-order ($M1$, $E2$) and hyperfine-induced $E1$ transitions modes \cite{tag2001, tbg2002,bja1998}. First measurements of the metastable hyperfine-induced decay rate in N$^{3+}$ was first performed at the Hubble Space Telescope with important implications to the isotopic abundance in an observed nebula \cite{bjp2002}. Values of  $E1M1$ decay rate in berylliumlike sulphur can also play an important role, specially because the majority of stable isotopes  ($^{32}$S and $^{34}$S) contains $I=0$  and have observable quantities in the solar coronae \cite{all1973, bja1998, pss2003}.  
 
Besides this astrophysical interest, there is also motivation for calculating the $E1M1$ two-photon decay mode coming from experiments aimed to test the standard model via the observation of parity nonconservation in berylliumlike uranium \cite{msg1996, msi1998}. Moreover, some subtle X-ray lines  coming from an electron cyclotron resonance (ECR) plasma might be attribute to charge state mechanisms involving the Be-like metastable state ${1s}^{2} 2s 2p$ $^3\mbox{P}_0$ \cite{gas2013}. 

There are no experimental results for the $E1M1$ decay rates in zero-spin berylliumlike ions available. Only recently, several dielectronic recombination resonances were clearly identified as coming from a parent ${1s}^{2} 2s 2p$ $^3\mbox{P}_0$ metastable state in xenon ($^{136}$Xe$^{+50}$ with $I=0$), which will lead to a forthcoming measurement of the respective $E1M1$ decay rate  \cite{bbk2012, bbh2015}. In these recently published works, the need of full relativistic calculations for this decay rate is emphasised.

\begin{figure}[tb]
\centering
\includegraphics[width=1.0\columnwidth]{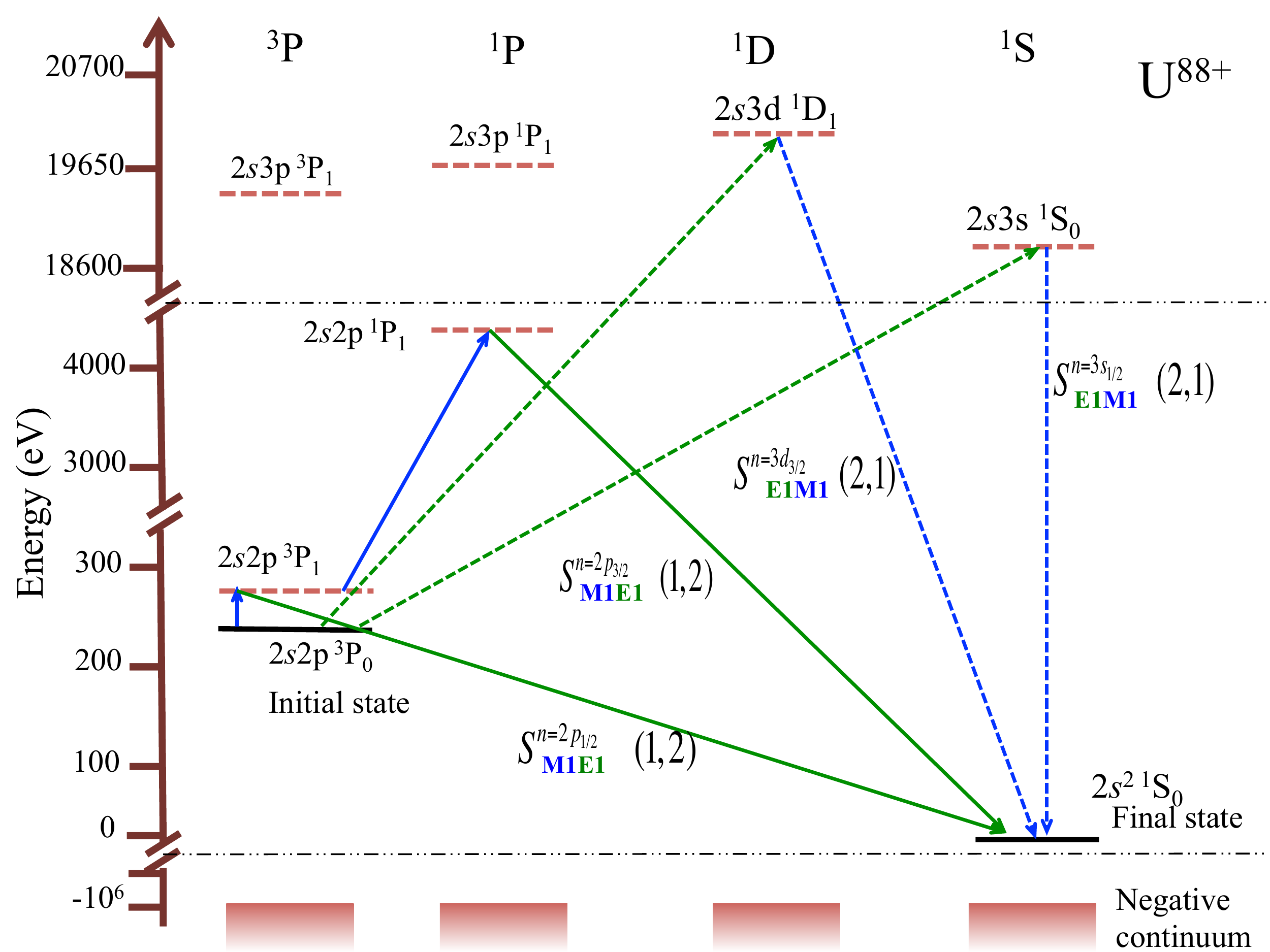}
    \caption{\label{fig:transit} (Color online) Energy atomic structure of berylliumlike uranium relative to the final state $2s^2~^1  \mbox{S}_0$. }
\end{figure}

 In isotopes with non-zero $I$ the importance of the $E1M1$ mode is reduced: the hyperfine mixing between the term $^3\mbox{P}_0$ and the closely-lying above term $^3\mbox{P}_1$ produces states with total angular momentum $F\ne0$, thus circumventing the $0\rightarrow0$ selection rule. This drastically reduces the lifetime of the metastable $1s^22s2p~^3\mbox{P}_0$ state since it opens an $E1$ single-photon channel.  Decay rates for this hyperfine-induced $E1$ mode have been known theoretically for some years \cite{mpi1993, bja1998, ccj2008, azh2009} and the first measurements have already been performed  recently, both in laboratory \cite{ssv2007, sbm2012} and in galaxy nebula \cite{bjp2002}. A review of this topic can be found in Ref.~\cite{joh2011}.

From the theoretical point of view, the calculation of this two-photon decay rate offers a challenge not only because berylliumlike ions have a compact electron structure, which makes electronic correlation of paramount importance, but also due to relativistic effects, such as the negative-energies. Previous studies about  two-photon decay with a $M1$ component  have shown that this negative-energy contribution is mandatory for both low-$Z$ and high-$Z$ ions \cite{lss2005, ssa2009} and improves gauge invariance \cite{sds1999a, ccj2001, isv2004}. Furthermore, similar investigations have concluded that the inclusion of negative-energy continuum gives better agreement with experimental data \cite{ind1996}.  Both effects have to be efficiently incorporated in the second-order summation over the intermediate states that characterizes  two-photon transitions.

Figure~\ref{fig:transit} illustrates the compact atomic structure in berylliumlike uranium, where the initial, intermediate and final states are plotted.

%
%

Up to now,  only two estimations of the $E1M1$ decay rate for berylliumlike systems are available \cite{sch1973, lau1980}, both assuming a non-relativistic approximation and using $LS$-coupling, which for high-$Z$ ions may lead to significant deviations.  Moreover, the summation over the intermediate states was only restricted to the first terms, $1s^2 2s 2p~^3\mbox{P}_1$ and $1s^2 2s 2p~^1\mbox{P}_1$.

%
%
In this work, we  calculate the two-photon decay rate of the metastable 1s$^2$2s2p~$^3$P$_0$ state in berylliumlike ions    considering a {\it relativistic} evaluation of the second-order summation in a $jj$-coupling active-electron. Negative energies are thus included and investigated. In order to take into account the electronic correlation, we perform the evaluation of the second-order summation via a finite-basis-set  and an effective local potential, with a few key intermediate states calculated using the MultiConfiguration Dirac-Fock (MCDF) method. For these evaluations, we consider xenon and uranium, following the reasoning above. For elements below xenon, we notice that the strong electronic correlation prevents the present method of retrieving a reliable  decay rate. A model beyond the active electron model is currently  under investigation. 

\section{Theory}
 The evaluation of two-photon related quantities have been discussed several times in the literature \cite{god1981, dej1997, spf2001}, we, therefore, present here only the final form suitable  for further discussion of the influence of the relativistic and electronic correlation effects.

   Two-photon processes are evaluated following a second-order perturbation theory, which overall contains a summation over the complete spectrum of a given Hamiltonian. Its elements are often referred as intermediate states.
   For the present case of the $E1M1$ two-photon decay between the states $1s^2 2s 2p$~$^3\mbox{P}_0$ and $1s^2 2s^2$~$^1\mbox{S}_0$ (terms are given for state identification), the differential decay rate is given by (atomic units),

 \begin{eqnarray}	
\frac{d W}{d\omega _{1}} &=&\frac{\omega _{1}\omega _{2}}{(2\pi
)^{3}c^{2}} %
\left| \sum_{j_{{n}}=1/2}^{3/2}  \left[
S^{j_{{n}}}(2,1)
+
S^{j_{{n}}}(1,2)       \right]
\right|^2 , \nonumber \\ 
\label{decaytwo_ele}
\end{eqnarray}
where $\omega_1$ and  $\omega_2$ are the energies of the two emitted photons, $c$ is the light speed and $j_n$ is the total angular momentum of the active electron performing the transition.  
From now on, we write configurations without $1s^2$  for shortness.
The sum of both photon energies is equal to transition energy due to energy conservation, $E_{2s 2p~^3\mbox{\tiny P}_0}- E_{2s^2~^1\mbox{\tiny S}_0}=\omega_1+\omega_2=\omega_t$.  
The two-photon amplitudes $S^{j_{{n}}}(2,1)$ and $S^{j_{{n}}}(1,2)$ contains the summation over the reduced matrix elements of the $E1$ and $M1$ multipole components, which are given by,
\begin{eqnarray}
S^{j_n}(2,1) =\hspace{5.25cm} \nonumber \\ 
{\sum_{  n }} 
\frac{\rmem{ 2s~^1\mbox{S}_0}{ R_2}{n_\nu j_\nu}
\rmem{n_\nu j_\nu}{ R_1}{  2p~^3\mbox{P}_0 }
}{E_{n} - E_{2s 2p \hspace{0.01 cm}^3\mbox{\small P}_0} + \omega_1}.
\label{eq:S_one_ele}
\end{eqnarray}
with the multipole components, electric dipole and magnetic dipole being given by the relativistic radiative operators $R_{1}=E_1$ and   $R_{2}=M_1$, respectively \cite{god1981}. $S^{j_n}(1,2)$ is given by an equation similar to Eq.~\eqref{eq:S_one_ele} by interchanging 1 with 2.
%
In Fig.~\ref{fig:transit} are represented the first states of the summation for the four two-photon amplitudes allowed by selection rules, which are $2s 3s_{1/2}~^1\mbox{S}_0$, $2s 3d_{3/2}~^1\mbox{D}_1$, $2s 2p_{1/2}~^3\mbox{P}_1$ and $2s 2p_{3/2}~^1\mbox{P}_1$, for $S^{1/2}(2,1)$, $S^{3/2}(2,1)$, $S^{1/2}(1,2)$ and $S^{3/2}(1,2)$, respectively. 

  In this work, we consider the \emph{active electron model} (AEM) \cite{spf2001}, i.e,  only intermediate states with variations of the active electron's quantum numbers $n'l'_{j'}$ that participates in the transition $2p\rightarrow n'l'_{j'}\rightarrow2s$ are taken into account in the summation over the intermediate states. Other intermediate states with excitation of the  spectator electron, (like the $1s$ and $2s$ occupied orbitals) are thus not taken into account. 
 Atomic states are usually given as a linear combination of configurations within a MCDF or configuration interaction (CI). We hereby define a state with major contribution of a configuration with a spectator-orbital excitation as $C^{\mbox{\scriptsize exc}}$. $C^{\mbox{\scriptsize non-exc}}$ are usual states within the AEM, where the major contribution addresses to non-excitation configurations of the spectator-orbital. 
\begin{figure}[tb]
\centering
\includegraphics[width=1.0\columnwidth]{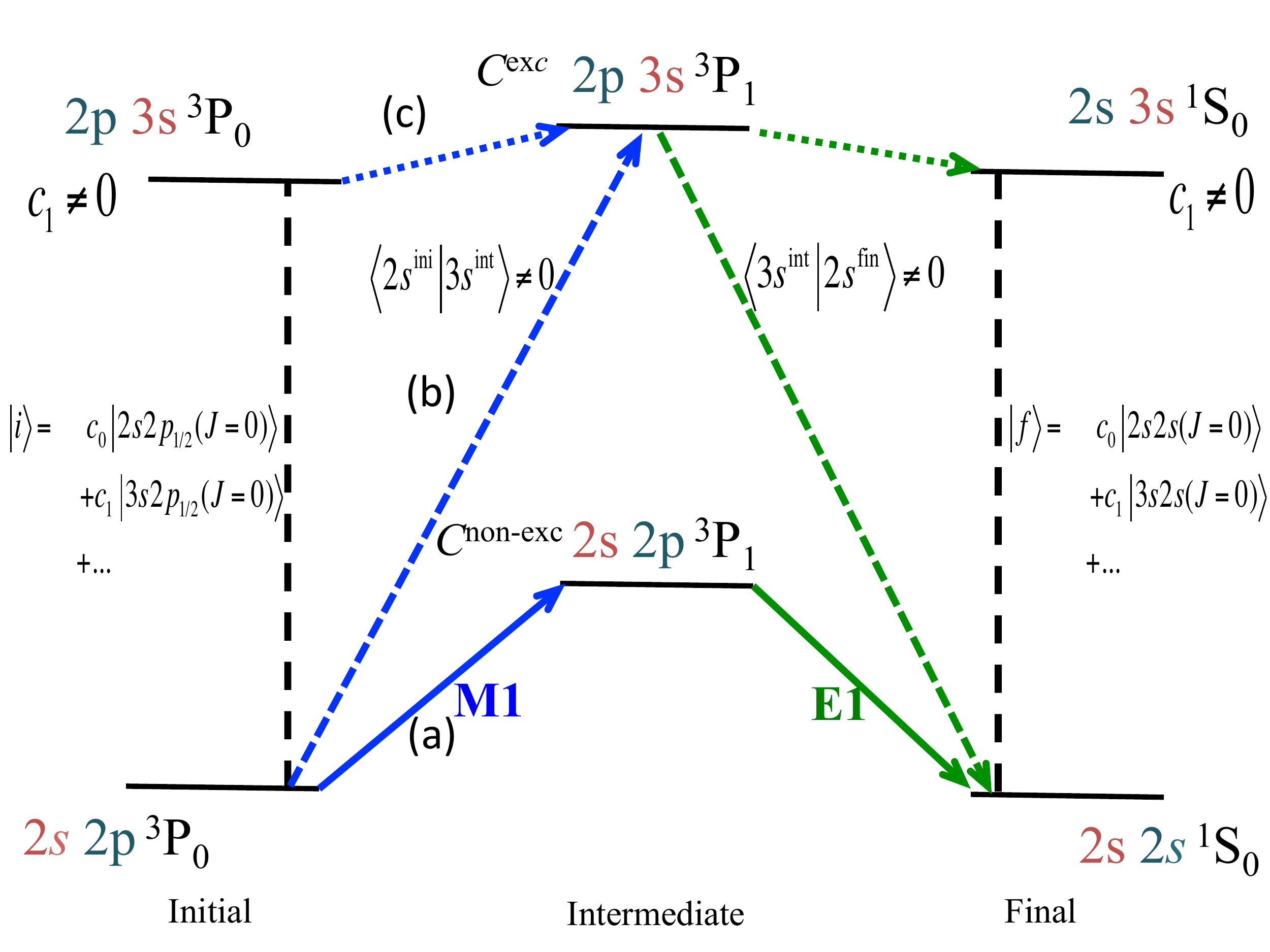}
    \caption{\label{fig:core_pic} (Color online) Representation of  $C^{\small \mbox{\scriptsize non-exc}}$ and $C^{\mbox{\scriptsize exc}}$ possible intermediate states.  The path (a) (solid arrows) corresponds to a intermediate state $C^{\mbox{\scriptsize non-exc}}$ within the AEM. The path (b) (dashed arrows) connects an intermediate state $C^{\mbox{\scriptsize exc}}$ to the final and initial states via the nonorthogonality of the spectator electrons. Path (c) (dot arrows) connects correlation configurations of the initial and final states to $C^{\mbox{\scriptsize exc}}$. The dashed line means a link between the configurations with ($c_0$) higher and ($c_1$) minor contributions in the $jj$-expansion.}
\end{figure}
 In order to better justify the AEM, we give in Figure \ref{fig:core_pic} a pictorial representation of one $C^{\mbox{\scriptsize exc}}$ and one $C^{\mbox{\scriptsize non-exc}}$.     
The path (a) corresponds to a state $C^{\mbox{\scriptsize non-exc}}$ in the AEM. This path is also represented in Fig.~\ref{fig:transit}. 
Because the radiative operator is a one-body operator, $C^{\mbox{\scriptsize exc}}$  states give non-null matrix elements only by considering either of these two cases: 
\begin{itemize} 
\item \emph{Path~(b)-} Electron orbitals are almost orthogonal between all states, thus if the radiative operator connects the active orbitals, there is a small contribution of $C^{\mbox{\scriptsize exc}}$ due to $\normv{2s^{\mbox{\scriptsize ini}}}{3s^{\mbox{\scriptsize int}}}\ne0$ and $\normv{3s^{\mbox{\scriptsize int}}}{2s^{\mbox{\scriptsize fin}}}\ne0$, where $2s^{\mbox{\scriptsize ini}}$,  $3s^{\mbox{\scriptsize int}}$ and $2s^{\mbox{\scriptsize fin}}$ are spectator orbitals in the initial, $C^{\mbox{\scriptsize exc}}$ and final states.
 \item \emph{Path~(c)-} The final and initial states can have a reasonable contribution of a configuration with the same spectator orbital as $C^{\mbox{\scriptsize exc}}$ due to configuration mixing. The  configuration coefficients can be obtained either by MCDF or CI.  
\end{itemize} 
These two cases show how multiconfiguration and  fully relaxed orbitals can play a role in two-photon processes by allowing $C^{\mbox{\scriptsize exc}}$  states beyond the AEM.  

 For the present elements of xenon and uranium, MCDF calculations addressed to the initial, intermediate and  final  states with fully relaxed orbitals shows data that justifies the use of the AEM:  First, due to the strong field of the nucleus, even for xenon, the obtained radial orbitals of the spectator electrons are reasonably orthogonal in all initial, final and intermediate states (residues of 2\%); Second, the $jj-$expansion of both initial, intermediate and final states are well represented by a single configuration (all other configuration coefficients adds up to 2\%).  
 
 For low-$Z$ ions or neutral beryllium, on the other hand, strong electron correlation does not allow the application of AEM. The intermediate state summation can be done either  via an inhomogeneous four-electron Dirac Hamiltonian (as in Ref.~\cite{dej1997} for heliumlike ions), or by the introduction of $C^{\mbox{\scriptsize exc}}$, which requires a careful analysis of the configuration mixing  coefficients and orthogonality.

In the present AEM, the evaluation of the two-photon amplitudes is performed by applying the finite-basis-set (FBS) method to the representation of the $C^{\mbox{\scriptsize non-exc}}$ intermediate states, which are eigenstates of the Dirac many-electron Hamiltonian. A  B-spline basis set \cite{sfi1998, asf2009} is considered  for  a cavity of radius 60 atomic units and 50 positive-energy and 50 negative energy states. Since the AEM is employed here, the FBS spectra addresses the active  electron. A degree of correlation is  introduced in order to match the respective orbitals obtained by the MCDF method. The local electrostatic potential formed by the $1s$ and $2s$ spectator orbitals, $2v_0(1s, r)+v_0(2s ,r) $ \cite{joh2007}, is considered in all active states, where $v_0(\nu, r)$ is given by
\begin{equation}
v_0(\nu, r)=\int \left(   P_\nu (r')^2 +  Q_\nu (r')^2 \right) \frac{1}{r_{>}} dr'~,
\label{eq:vo}
\end{equation}
Here, $P_\nu$ and $Q_\nu$ are the large and small components of the radial wavefunctions of a spectator orbital $\nu$ and $r_{>}=\mbox{max}(r,r')$.
A comparison of the spectator orbitals of all states obtained by MCDF shows 5\% differences, which for the evaluation of the electrostatic potential can be neglected. The spectator orbitals of the $2s2p~^3\mbox{P}_0$ state are chosen for $v_0(1s, r)$ and $v_0(2s, r)$. A local statistical-exchange potential is also included in order to approximate the non-local part of the Dirac-Fock equation. We follow the original procedure of  Cowan \cite{cow1967} that defines this local potential for an orbital $\nu$ as, 
\begin{eqnarray}
v_{ \mbox{\tiny exc}} (\nu, r)&=&-k_1 \phi(r)  \left[  \frac{\rho' (r)}{\rho' (r) + 0.5/(n_\nu -l_\nu)}  \right]  \nonumber \\
&\times&\left( \frac{\rho'(r)}{\rho(r)} \right) \left( \frac{ 24 \rho(r)}{\pi} \right)^{1/3}~,  
\label{eq:vexc}
\end{eqnarray}
where $\rho$ is the  many-electron total electron density and $\rho' (r)$ is the modified total  density without the contribution of the $\nu$  orbital, i.e., $\rho' (r)=\rho(r)-\mbox{min}(2, e_\nu)\rho_{\nu}(r)$, with $\rho_{\nu}(r)$ being the electron density of the orbital $\nu$.  The quantity $e_\nu$ is the number of equivalent electrons at the orbital $\nu$ with principal quantum number and orbital angular mometum, $n_\nu$ and $l_\nu$, respectively. The function $\phi(r)$ takes into account the different influence of the centrifugal potential to the various orbitals as described in Ref.~\cite{cow1967}. All the present wavefunctions and densities necessary for calculating Eqs.~\eqref{eq:vo} and \eqref{eq:vexc} were obtained by the MCDF method. 

\begin{table}[t]
\caption{\label{tab:k1_energi} Optimal values of $k_1$ and  energy differences (eV) obtained by the FBS method  without $k_1$ optimization ($E_{\mbox{\tiny FBS}}^{*}$), as well as the respective ones obtained by the MCDF method ($E_{\mbox{\tiny MCDF}}$).  Energy of 2s2p $^3$P$_0$ is relative to final state 2s$^2$ $^1$S$_0$ ($\omega_t$), while the rest are relative to the initial state, $E_{n} - E_{2s 2p_{1/2}~^3\mbox{\tiny P}_0}$. Values provided by Ref.~\cite{sjs1996} are also listed.
} 
\begin{ruledtabular}
\begin{tabular}{llD{.}{.}{3}D{.}{.}{3}D{.}{.}{3}D{.}{.}{3}}

&    &   \multicolumn{1}{l}{  $k_1$ }  &     \multicolumn{1}{l}{ $E_{\tiny \mbox{FBS}}^{*}$  } &   \multicolumn{1}{l}{ $E_{\tiny \mbox{MCDF}}$ } &   \multicolumn{1}{l}{ Ref.~\cite{sjs1996} }  \\
%
Xe$^{50+}$ &  $^3$P$_0-^1$S$_0$ &0.79	&	118.5	&	104.1	&	104.5	\\
		   &  $^3$P$_1-^3$P$_0$ & 0.64	&	0.0	&	23.9	&	22.8	\\
		  &   $^1$P$_1-^3$P$_0$ &0.62	&	400.8	&	430.1	&	428.3	\\	             
	     \\[-2.0ex]        \hline	 \\[-2.0ex] 	     
U$^{88+}$  & $^3$P$_0-^1$S$_0$  &0.69	&	252.7	&	258.1	&	258.3	\\
	             & $^3$P$_1-^3$P$_0$& 0.58	&	0.0	&	41.6	&	39.9	\\
	             &  $^1$P$_1-^3$P$_0$ &0.72	&	4259.3	&	4245.3	&	4243.3	\\
\end{tabular}
\end{ruledtabular}
\end{table}
Next,  we identify the intermediate states with the most relevant weight to the summations  and calculate their most accurate MCDF energies $E_n$. These intermediate states are depicted in Fig.~\ref{fig:transit}.
 While the parameter $k_1$ is set to 0.7 in Ref.~\cite{cow1967} as the best empirical guess for the exchange potential, we here consider it as a free parameter. Optimal values of this parameter are obtained by comparing the values of the transition energy ($\omega_t$) and the  energy differences $E_{n} - E_{2s 2p_{1/2}~^3\mbox{\tiny P}_0}$ (denominators of Eq.~\eqref{eq:S_one_ele}), obtained by the FBS method and with the respective ones of the MCDF method. 
Table~\ref{tab:k1_energi} lists the optimal values of $k_1$ that minimizes the differences between the FBS and MCDF of the mentioned energy differences.  
 The  MCDF calculations were performed using the general relativistic MCDF code (MDFGME) \cite{ind1990}.  
 
Calculations of the decay rate were performed in both length and velocity gauges. The quality of the evaluation of the two-photon amplitudes, if the potential remains local in all states, is directly connected to the gauge invariance \cite{god1981, asf2009}.   Although we introduced different local-exchange potentials in the states and MCDF  energies, we notice that the gauge invariance is still at a level of few percent. 

With the application of the present formalism to the decay of 1s2p~$^3P_0$ to the ground state in heliumlike ions, and with an effective potential of $v_0(1s,r)$, we reproduce the results of Ref.~\cite{saj2002} within the respective accuracy.

\section{Results and Discussion}
\label{sec:res_dis}

\begin{table}[b]
\caption{\label{tab:decays}
Decay rate (s$^{-1}$) for $2s2p~^3\mbox{P}_0\rightarrow2s^2~^1\mbox{S}_0$ $E1M1$ transition in xenon and  uranium. Relativistic calculations have been performed in  velocity (V) and length (L) gauges for several cases: with ($W^{\mbox{\scriptsize opt}}$) and without $k_1$-optimization ($W^{\mbox{\scriptsize non-opt}}$);  with the summation carried without negative energies ($W^+$); having the energies provided by Ref.~\cite{sjs1996} ($W^{*}$); without the effective exchange potential ($W^{\mbox{\scriptsize no-exc}}$). Values of Refs~\cite{lau1980} and \cite{bbk2012} are listed. }
%
\begin{ruledtabular}
\begin{tabular}{lllll}
 &   \multicolumn{1}{c}{ $W_{\mbox{\scriptsize V}}^{\mbox{\scriptsize opt}}$ }  &   \multicolumn{1}{c}{ $W_{\mbox{\scriptsize L}}^{\mbox{\scriptsize opt}}$ } &  
    \multicolumn{1}{c}{ $W_{\mbox{\scriptsize V}}^{\mbox{\scriptsize non-opt}}$ } &   \multicolumn{1}{c}{ $W_{\mbox{\scriptsize L}}^{\mbox{\scriptsize non-opt}}$ } \\
     \\[-2.0ex] 
   Xe$^{50+}$     &	$4.78  \times10^{-3}$ & $ 4.97\times10^{-3}  $  &   $4.05\times10^{-3}$&  $4.05\times10^{-3}$	\\
   U$^{88+}$ 	&	$8.04 \times10^{-2}$& $ 8.06\times10^{-2}  $    & $9.35\times10^{-2}$    & $9.35\times10^{-2}$ \\
\hline
 \\[-2.0ex] 
& \multicolumn{1}{c}{$W_{\mbox{\scriptsize V}}^+$} &  \multicolumn{1}{c}{$W_{\mbox{\scriptsize L}}^+$} &  
    \multicolumn{1}{c}{ $W_{\mbox{\scriptsize V}}^{*}$  } &   \multicolumn{1}{c}{ $W_{\mbox{\scriptsize L}}^{*}$ } \\
     \\[-2.0ex] 
\hline
\\[-2.0ex] 
Xe$^{50+}$     &	$4.78 \times10^{-3}$ & $ 4.98\times10^{-3}  $ &  $5.20\times10^{-3}$&  $5.40\times10^{-3}$	\\
 U$^{88+}$ 	&	$8.08  \times10^{-2}$& $ 8.11\times10^{-2}  $	& $8.18\times10^{-2}$    & $8.20\times10^{-2}$ \\
\\[-2.0ex]  \cline{2-5} \\[-2.0ex]	\\[-2.0ex]	
 &   \multicolumn{1}{c}{ $W^{\mbox{\scriptsize no-exc}}$ }  &   Ref.~\cite{lau1980}& Ref.~\cite{bbk2012}\footnote{Extension of Ref.~\cite{lau1980} having the energy splitting  $^3\mbox{P}_0-^3\mbox{P}_1$ into account.} \\
Xe$^{50+}$      & $ 5.30\times10^{-3}  $ &  $3.4\times10^{-2}$&  $5.2\times10^{-2}$	\\
 U$^{88+}$ 	& $ 8.31\times10^{-2}  $	& $2.6\times10^{ 1}$    &   $4.9\times10^{1}$  \\

\end{tabular}
\end{ruledtabular}
\end{table}

The results of our calculations for the $2s 2p~^3\mbox{P}_0\rightarrow 2s^2~^1\mbox{S}_0$ $E1M1$ decay rate $W$   are presented in Table~\ref{tab:decays}. The obtained lifetimes corresponds to $\sim$3 min and 12~s for Xe$^{50+}$ and  U$^{88+}$ ions, respectively.  Other allowed higher-order multipole contributions to this transition, like the $E2M2$ or $E3M3$, are severely reduced. The obtained value for the $E2M2$ decay rate in beryliumlike uranium of 8.2$\times10^{-18}$~s$^{-1}$ shows the minimal impact to the total decay rate.

Calculations were performed in both velocity and length gauges, showing differences of up to 4\% due to the different local-exchange potentials in the states. The case without these effective exchange potentials ($W^{\mbox{\scriptsize no-exc}}$) results in a gauge invariance of 10$^{-10}$\%.

Differences between the values of the decay rate with and without  $k_1$-optimization in Table~\ref{tab:decays}  are mostly due to the respective transition energies, for which the decay rate depends quadratically, as well on the different $^3$P$_1$ and $^1$P$_1$ energies. These values can also can be compared with the case of not considering the effective exchange potential of Eq.~\eqref{eq:vexc}. Differences of up to  20\% and 16\% in xenon and uranium, respectively, shows how sensitive the decay of this transition is to the electronic correlation, in particular to the non-local part of the electron-electron interaction. 

 Residual differences of $0.5-2$~eV  between MCDF energy values and those of Ref.~\cite{sjs1996}  results in relative differences of   8\% in the decay rate.  Most of the experimental observations \cite{fzv2005, bkm1993, bbh2015} and theoretical calculations \cite{sjs1996, chc1997b, smf1998, saf2000,ccs2000} of these energies are included in a energy range of 2~eV, resulting in differences up to 10\%. 
 
 To be conservative, we consider the uncertainty in the decay rate as the combined uncertainty of the previous  effects. The final result of the $E1M1$ decay rate is thus equal to $(5\pm1)\times10^{-3}$ s$^{-1}$ and $(8\pm1)\times10^{-2}$ s$^{-1}$ for xenon and uranium, respectively. 
 
 In contrast to previous studies of the negative continuum, where it shows that its contribution has to be included in relativistic calculations of two-photon decay rates \cite{lss2005, ssa2009}, the present case is of order of few percent, even for berylliumlike uranium ions. Following the semirelativistic approach of Ref.~\cite{saj2002}, the estimation of the negative-continuum contribution to the decay is proportional to $\omega_t^5/Z^2$. Previous studies deal with transitions between principal quantum numbers (e.g. \cite{ssa2009}), for which the transitions energies scales as $Z^4$. In the present case, the transition addresses the same quantum number and scales roughly as $Z$. This makes a smaller contribution of the negative energy continuum.

     \begin{figure}[tb]
\centering
\includegraphics[width=1.0\columnwidth]{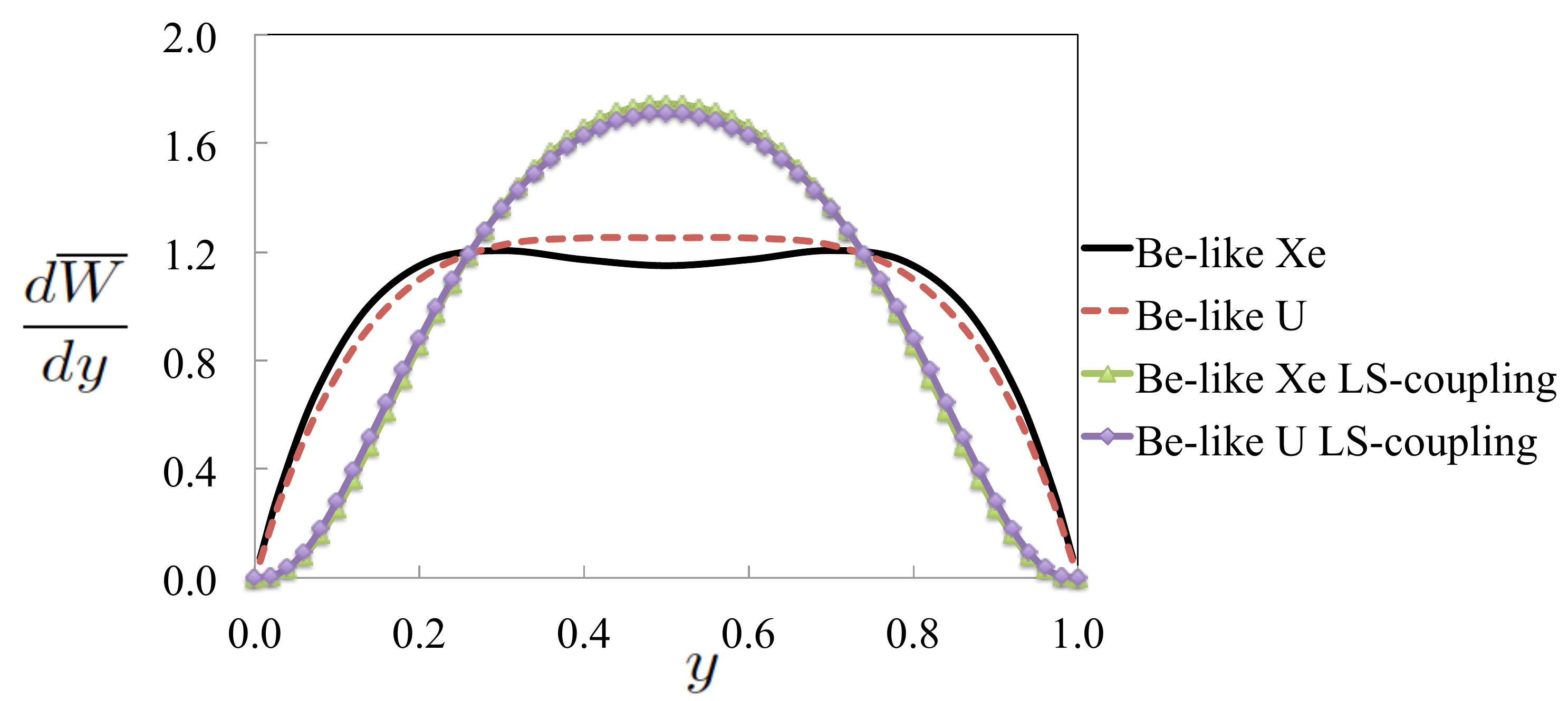}
    \caption{\label{fig:diff_dec} (Color online) Normalized differential decay rate for Xe$^{50+}$ and U$^{88+}$.  Values in a $LS$-coupling scheme are from Ref.~\cite{sch1973}. The quantity $y$ represented the energy sharing between both photons, i.e., $y=\omega_1/\omega_t$. Fx figure} 
\end{figure}

We notice evident differences relative to previous calculations by factors from 10 to 300.  The differences can be attributed to our full relativistic approach in a $jj$-coupling scheme.  This can be further investigated in the differential decay rate that is illustrated in Fig.~\ref{fig:diff_dec}, where it is shown normalized values (to the integral) of Eq.~\eqref{decaytwo_ele}. 
Here, values for Xe$^{50+}$ and U$^{88+}$ ions obtained in this work and by Ref.~\cite{sch1973} are displayed, which show evident differences in the differential decay rate. 

The values of Refs.~\cite{sch1973, lau1980}  where obtained by considering only the 2s2p$^3P_1$ and  2s2p$^1P_1$ states in the intermediate-state summation and were calculated in a non-relativistic $LS$-coupling framework. Moreover, the non-relativistic form of the electric and magnetic dipole operators was also employed in Refs.~\cite{sch1973,lau1980}, which forbids intercombination transitions with a spin-flip of the total spin in a $LS$-coupling. Therefore, spin-orbit and spin-spin interactions were included in first approximation in order to mix the $^3P_1$ and  $^1P_1$ terms.  For highly charged ions, intercombination transitions are allowed in a $jj$-coupling scheme with relativistic wavefunctions, as the spin-orbit interaction is already included non-perturbately.
 Other investigations of the $E1E1$ have shown that relativistic effects increase the decay rate by 30\% \cite{dra1986, dej1997} in heliumlike Xe.  
In the present case, the $M1$ mode is even more sensitive  to the $LS$-coupling scheme that is not appropriate for highly charged ions, where the strong spin-orbit interaction is included perturbately. 
A similar factor of 300 was already obtained in a relativistic calculation \footnote{Fratini,  private report (2013).}.

\section{Conclusion}  
\label{sec:sum}

We have presented the results of the
two-photon forbidden $E1M1$ decay rate for two selected heavy elements obtained with an effective potential. The limitations of the active electron model for this particular decay is investigated and found that while this approach cannot be applied to low- and middle-$Z$ ions, for berylliumlike Xe and heavier elements, each state is well described by a single configuration with orthogonal orbitals. Therefore, excitations of the spectator electron that forbids the use of this model can be neglected. We have
found a negligible contribution of negative-energy states to
 this decay rate, which is in agreement with semirelativistic estimations. 
On the other hand, we observe significant relativistic effects relative to non-relativistic calculations performed for middle-$Z$  ions, which can be attributed to the fact that the $LS$-coupling scheme  is not appropriate for the evaluation of this decay rate in highly charged ions.

%
\begin{acknowledgments}

This research was supported in part by Funda\c{c}\~{a}o para a Ci\^{e}ncia e a  Tecnologia (FCT), Portugal,
through the project No.  \emph{PTDC/FIS/117606/2010}, financed by the European
Community  Fund FEDER through the COMPETE.
P.~A., J.~M. and M.~G. acknowledge the support of the FCT, under Contracts No. \emph{SFRH/BPD/92329/2013} and \emph{SFRH/BD/52332/2013}, \emph{SFRH/BPD/92455/2013}.    
F.F. acknowledges support by the Austrian Science Fund (FWF) through the START grant \emph{Y 591-N16}. 
L.S. acknowledges
financial support from the People Programme (Marie Curie Actions) of the European Union's Seventh Framework Programme (\emph{FP7/2007-2013}) under REA Grant Agreement No. [\emph{291734}]. 

\end{acknowledgments}



\bibliography{TPBE_articles}

\end{document}